\documentclass[aps,prl,twoside,twocolumn,superscriptaddress,floatfix,nofootinbib,showpacs]{revtex4}
\usepackage{amsmath}
\usepackage{bm}
\usepackage{xcolor}

\newcommand\nhat{\hat{\mathbf n}}

\newcommand\beq{\begin{equation}}
\newcommand\eeq{\end{equation}}
\newcommand\beqn{\begin{eqnarray}}
\newcommand\eeqn{\end{eqnarray}}

\newcommand\fsky{f_{\mathrm{sky}}}

\newcommand{\ba}{\begin{eqnarray}}
\newcommand{\ea}{\end{eqnarray}}
\newcommand{\be}{\begin{equation}}
\newcommand{\ee}{\end{equation}}

\newcommand\lsim{\mathrel{\rlap{\lower4pt\hbox{\hskip1pt$\sim$}}
        \raise1pt\hbox{$<$}}}
\newcommand\gsim{\mathrel{\rlap{\lower4pt\hbox{\hskip1pt$\sim$}}
        \raise1pt\hbox{$>$}}}

\newcommand{\jcap}{{J.~Cosm.~Astrop.~Phys.}}
\newcommand{\araa}{{Annu.~Rev.~Astron.~Astrophys.}}
\newcommand{\aap}{{Astron.~Astrophys.}}
\newcommand{\apjl}{{Astrophys.~J.~Lett.}}
\newcommand{\apjs}{{Astrophys.~J.~Supp.}}

\newcommand{\mnras}{{Mon.~Not.~R.~Astron.~Soc.}}
\usepackage{graphicx}
\usepackage[large]{subfigure}
\usepackage{amssymb, amsmath}
\usepackage[amssymb]{SIunits}
\usepackage{epstopdf}
\usepackage{aas_macros}
\usepackage{natbib}

\begin{document}

\title{Taking the Universe's Temperature with Spectral Distortions of the Cosmic Microwave Background}
%%%%%%%%%%%%%%%%%%%%%%%%%%%%%%%%%%%%%%%%%%%%%%%%%%%%%%%%%%%%%%%%%%%%%%%%%%%
 \author{J.~Colin~Hill}\email{jch@astro.columbia.edu}
 \affiliation{Dept.~of Astronomy, Pupin Hall, Columbia University, New York, NY USA 10027}
 \author{Nick~Battaglia}
 \affiliation{Dept.~of Astrophysical Sciences, Peyton Hall, Princeton University, Princeton, NJ USA 08544}
  \author{Jens~Chluba}
 \affiliation{Kavli Institute for Cosmology, University of Cambridge, Cambridge, UK CB3 0HA}
  \author{Simone~Ferraro}
 \affiliation{Dept.~of Astrophysical Sciences, Peyton Hall, Princeton University, Princeton, NJ USA 08544}
  \author{Emmanuel~Schaan}
 \affiliation{Dept.~of Astrophysical Sciences, Peyton Hall, Princeton University, Princeton, NJ USA 08544}
  \author{David~N.~Spergel}
 \affiliation{Dept.~of Astrophysical Sciences, Peyton Hall, Princeton University, Princeton, NJ USA 08544}
%%%%%%%%%%%%%%%%%%%%%%%%%%%%%%%%%%%%%%%%%%%%%%%%%%%%%%%%%%%%%%%%%%%%%%%%%%%
\begin{abstract}
The cosmic microwave background (CMB) energy spectrum is a near-perfect blackbody.  The standard model of cosmology predicts small spectral distortions to this form, but no such distortion of the sky-averaged CMB spectrum has yet been measured.  We calculate the largest expected distortion, which arises from the inverse Compton scattering of CMB photons off hot, free electrons, known as the thermal Sunyaev-Zel'dovich (tSZ) effect.  We show that the predicted signal is roughly one order of magnitude below the current bound from the \emph{COBE-FIRAS} experiment, but can be detected at enormous significance ($\gtrsim 1000\sigma$) by the proposed \emph{Primordial Inflation Explorer} (\emph{PIXIE}). Although cosmic variance reduces the effective signal-to-noise to $230\sigma$, this measurement will still yield a sub-percent constraint on the total thermal energy of electrons in the observable universe.  Furthermore, we show that \emph{PIXIE} can detect subtle relativistic effects in the sky-averaged tSZ signal at $30\sigma$, which directly probe moments of the optical depth-weighted intracluster medium electron temperature distribution.  These effects break the degeneracy between the electron density and temperature in the mean tSZ signal, allowing a direct inference of the mean baryon density at low redshift.  Future spectral distortion probes will thus determine the global thermodynamic properties of ionized gas in the universe with unprecedented precision.  These measurements will impose a fundamental ``integral constraint'' on models of galaxy formation and the injection of feedback energy over cosmic time.
\end{abstract}
\maketitle

\emph{Introduction---}
The energy spectrum of the cosmic microwave background (CMB) radiation is extremely close to a perfect blackbody~\cite{Mather1994,Fixsen1996}.  However, distortions to this blackbody spectrum arise from physical processes in both the early (redshift $z > 1100$) and late ($z < 1100$) universe.  These processes include energy injection from decaying or annihilating particles~(e.g.,~\cite{Hu-Silk1993,McDonald2001}), the dissipation of small-scale primordial density fluctuations~(e.g.,~\cite{Daly1991,Barrow-Coles1991,Chluba2012}), and more exotic processes, such as energy injection from cosmic strings~\cite{Ostriker-Thompson1987} or primordial black holes~\cite{Carr2010}.  The most well-understood distortion is that due to the thermal Sunyaev-Zel'dovich (tSZ) effect, which is the inverse Compton scattering of CMB photons off hot, free electrons~\cite{Zeldovich-Sunyaev1969,Sunyaev-Zeldovich1970}.  This signal is predominantly due to hot gas in the intracluster medium (ICM) of galaxy groups and clusters at $z \lesssim 2$.

Many analyses have constrained cosmology and ICM physics using tSZ measurements of cluster samples or the fluctuation properties of the tSZ field~(e.g.,~\cite{Hasselfieldetal2013,Reichardtetal2013,Planck2015clusters,Planck2015ymap,Hilletal2014,Crawfordetal2014}).  However, the mean tSZ signal of the universe has yet to be detected --- in fact, no global spectral distortion of the CMB has been seen to date.  The tightest constraint comes from the \emph{COBE-FIRAS} experiment, which found $|\langle y \rangle| < 1.5 \times 10^{-5}$ at 95\% confidence~\cite{Fixsen1996}, where $y$ is the Compton-$y$ parameter characterizing the (non-relativistic) tSZ distortion,
\be
\label{eq.ydef}
y(\nhat) = \frac{\sigma_{\rm T}}{m_{\rm e} c^2} \int dl \, n_{\rm e}(\nhat,l) k_{\rm B} T_{\rm e}(\nhat,l) \,.
\ee
Here, $n_{\rm e}$ and $T_{\rm e}$ are the electron number density and temperature, respectively, $\sigma_{\rm T}$ is the Thomson cross-section, $m_{\rm e} c^2$ is the electron rest-mass energy, and $k_{\rm B}$ is Boltzmann's constant.  This expression assumes $T_{\rm e} \gg T_{\rm CMB}$, which is valid for all sources in our analysis.  Here, $T_{\rm CMB} = 2.726 \pm 0.001$ K is the average CMB temperature today~\cite{Fixsen2009}.  The integral is taken along the line-of-sight distance $l$, while $\nhat$ is a position on the sky.  Angle brackets $\langle \cdots \rangle$ refer to averages taken over the sky.  We show below that the expected $\langle y \rangle$ signal is roughly one order of magnitude below the \emph{COBE-FIRAS} limit and will be easily detectable with currently proposed experiments.

Recently, interest has grown in new measurements of the CMB energy spectrum~\cite{PIXIE2011,PRISM2014}.  Such measurements have the potential to detect or place interesting constraints on a range of new physics, such as primordial non-Gaussianity~\cite{Pajer-Zaldarriaga2012} or decaying dark matter~\cite{Chluba-Sunyaev2012,Chluba2013}.  The CMB specific intensity, $I_{\nu}^{\rm CMB}$, at frequency $\nu$ in a direction on the sky, $\nhat$, can be decomposed into:
\ba
\label{eq.DeltaIdef}
I_{\nu}^{\rm CMB}(\nhat) & = & B_{\nu}(T_{\rm CMB}+\Delta T) + \Delta I_{\nu}^T(\nhat) + \Delta I_{\nu}^{\rm tSZ}(\nhat) + \nonumber \\
 & & \Delta I_{\nu}^{\mu}(\nhat) + \Delta I_{\nu}^{\rm other}(\nhat) \,,
\ea
where $B_{\nu}(T) = \frac{2 h \nu^3}{c^2} / (e^x - 1)$ with $x = h \nu / (k_{\rm B} T)$.  The focus of this paper is the sky-averaged tSZ signal, $\langle \Delta I_{\nu}^{\rm tSZ} \rangle$.  %For the purposes of estimating component-separation noise, we compute the other terms in Eq.~(\ref{eq.DeltaIdef}) as follows.
Since the exact value of the primordial CMB temperature is not known to the nK level accessible to future experiments, it must be determined in the analysis, which is represented by the $\Delta T$ term in Eq.~(\ref{eq.DeltaIdef}). We assume a fiducial value of $\Delta T = 1.2 \times 10^{-4}$ K, within the $1\sigma$ error from \emph{COBE-FIRAS}.  The $\Delta I_{\nu}^{T}(\nhat)$ term in Eq.~(\ref{eq.DeltaIdef}) represents anisotropies in the primordial CMB temperature, which in a given pixel are indistinguishable from $\Delta T$.  The other terms in Eq.~(\ref{eq.DeltaIdef}) represent contributions from the $\mu$-distortion ($\Delta I_{\nu}^{\mu}$) and all other contributions to the sky intensity ($\Delta I_{\nu}^{\rm other}$), such as other sources of CMB spectral distortions or foregrounds (e.g., dust or synchrotron emission).  For the $\mu$-distortion, we assume a constant amplitude of $\mu = 2 \times 10^{-8}$, consistent with predictions from the damping of small-scale modes in the concordance $\Lambda$CDM cosmology~\cite{Chlubaetal2012}.  We set $\Delta I_{\nu}^{\rm other} = 0$.  Under these assumptions, we examine detection prospects for $\langle \Delta I_{\nu}^{\rm tSZ} \rangle$ for upcoming experiments, specifically the proposed \emph{Primordial Inflation Explorer} (\emph{PIXIE})~\cite{PIXIE2011}.

We show that \emph{PIXIE} can make a high-significance measurement of not only~$\langle y \rangle$, but also the mean optical depth-weighted ICM electron temperature, $\langle k_{\rm B} T_{\rm e} \rangle_{\tau}$, where $\tau (\nhat) = \sigma_{\rm T} \int dl \, n_{\rm e}(\nhat,l)$.  The $\langle k_{\rm B} T_{\rm e} \rangle_{\tau}$ sensitivity arises from the relativistic tSZ signal due to the hot ICM of galaxy groups and clusters, and thus calibrates the amplitude of the cluster mass -- temperature scaling relation that will be fundamental for cosmological constraints from X-ray cluster counts in the upcoming \emph{eROSITA} mission~\cite{Merloni2012}.  Measurements of $\langle y \rangle$ and $\langle k_{\rm B} T_{\rm e} \rangle_{\tau}$ provide ``integral constraints'' on models of galaxy formation over cosmic time, similar to the constraint on the optical depth to reionization, $\tau_{\rm CMB}$, provided by large-scale CMB polarization measurements.  CMB spectral distortion signals thus constrain the uncertain but crucial feedback mechanisms used in hydrodynamical simulations of cosmic structure formation~(e.g.,~\cite{Vogelsberger2014,Hopkins2014,Schaye2015}).  Moreover, constraints on the electron temperature through the relativistic tSZ effects allow an inference of the mean electron density from $\langle y \rangle$, yielding a definitive answer to the long-standing ``missing baryons'' problem at low redshift~(e.g.,~\cite{Cen-Ostriker1999,Fukugita-Peebles2004,Bregman2007}).  In a companion paper~\cite{B15}~(hereafter B15), we provide technical details of our model and calculations, and combine the forecasted measurements of $\langle y \rangle$ and $\langle k_{\rm B} T_{\rm e} \rangle_{\tau}$ with other data to constrain cosmological parameters and ICM gas physics models.

\emph{Theory---}
The sky-averaged tSZ signal receives contributions from three main components: the ICM of collapsed halos (galaxies, groups, and clusters), the intergalactic medium (IGM) between halos, and the epoch of reionization.  The latter two contributions are subdominant to that from the ICM, but all are included in our analysis.  Physically, the global tSZ signal probes the injection of energy into ionized gas over the history of the universe.  Using Eq.~(\ref{eq.ydef}),
\be
\langle y \rangle = \frac{\sigma_{\rm T}}{m_{\rm e} c^2} \int \frac{d^2 \nhat}{4\pi} \int dl P_{\rm e}(\nhat, l) \,\, \propto E_{\rm e}^{\rm th, tot} \,,%\nonumber \\
%   & \propto & (E_{\rm e}^{\rm coll} - E_{\rm e}^{\rm cool} + E_{\rm e}^{\rm inj}) \,,
\label{eq.Etot}
\ee
where the integrals are taken over the observable universe, $P_{\rm e} = n_{\rm e} k_{\rm B} T_{\rm e}$ is the electron pressure, and $E_{\rm e}^{\rm th, tot} = E_{\rm e}^{\rm coll} - E_{\rm e}^{\rm cool} + E_{\rm e}^{\rm inj}$ is the total thermal energy in electrons. Here, $E_{\rm e}^{\rm coll}$ is the energy in electrons due to gravitational collapse during structure formation, $E_{\rm e}^{\rm cool}$ is the electron energy lost to cooling processes, and $E_{\rm e}^{\rm inj}$ is the energy injected into electrons by feedback processes (e.g., from supernovae and active galactic nuclei).  For sufficiently high electron temperatures ($k_{\rm B} T_{\rm e} / m_{\rm e} c^2 \gtrsim 10^{-2}$), relativistic corrections to the tSZ effect become important, inducing an additional dependence on $T_{\rm e}$~\cite{Challinor-Lasenby1998,Nozawaetal2006,ChlubaSwitzer2013}.

The ICM and IGM contributions to $\langle \Delta I_{\nu}^{\rm tSZ} \rangle$ can be calculated directly from hydrodynamical simulations~\cite{Refregieretal2000,daSilvaetal2000,Seljaketal2001,Battagliaetal2010}.  The ICM signal can also be calculated in the halo model~(e.g.,~\cite{Cooray-Sheth2002}) using calibrated prescriptions for the electron pressure and temperature as a function of halo mass and redshift~\cite{Barbosaetal1996,Refregieretal2000}.  We define the IGM as gas located at greater than 2.5 times the virial radius of any collapsed halo.%We provide the main results here and defer full details to B15.

The sky-averaged tSZ signal from ICM electrons is (e.g., Eq.~(2) of Ref.~\cite{Hill-Sherwin2013})
\be
\label{eq.meantSZ}
\langle \Delta I_{\nu}^{\rm tSZ} \rangle = \int dz \frac{d^2 V}{dz d\Omega} \int dM \frac{dn}{dM} \int d^2 \nhat \, \Delta I^{\rm tSZ}_{\nu}(\nhat,M,z) \,,
\ee
where $d^2 V/dz d\Omega$ is the comoving volume per steradian at redshift $z$, $dn/dM$ is the halo mass function (the comoving number density of halos as a function of mass $M$ and redshift), and $\Delta I^{\rm tSZ}(\nhat,M,z)$ is the tSZ signal at angular position $\nhat$ on the sky with respect to the center of a cluster of mass $M$ at redshift $z$.  In the non-relativistic approximation,
\be
\label{eq.tSZnonrel}
\Delta I^{\rm tSZ, non-rel.}_{\nu}(\nhat,M,z) = g_{\nu} y(\nhat,M,z) \,,
\ee
where the tSZ spectral function (in intensity units) is
\be
\label{eq.gfunc}
g_{\nu} = \frac{2 \left(k_{\rm B} T_{\rm CMB}\right)^3}{\left(hc\right)^2} \frac{x^4 e^{x}}{\left(e^{x} - 1\right)^2} \left[ x \coth\left( \frac{x}{2} \right) - 4 \right] \,,
\ee
with $x = h \nu / (k_{\rm B} T_{\rm CMB})$.  In the relativistic case, corrections to the spectral function arise that explicitly depend on the electron temperature:
\be
\label{eq.tSZrel}
\Delta I^{\rm tSZ, rel.}_{\nu}(\nhat,M,z) = g_{\nu} \left[ 1+\delta^{\rm rel.}_{\nu}(\nhat,T_{\rm e}) \right] y(\nhat,M,z) \,.
\ee
We include relativistic corrections up to third order in $k_{\rm B} T_{\rm e}/(m_{\rm e} c^2)$ following Ref.~\cite{Nozawaetal2006}.  Note that relativistic corrections involving the cluster velocity also appear in general, but when averaged over all clusters in the universe, these cancel to lowest order due to the differing sign of the line-of-sight velocity from cluster to cluster.  The neglected, next-order contribution corresponds to $\langle y \rangle \approx 10^{-8}$~\cite{Hu1994} and its minor effect is left to future work. %The lowest order neglected term is O(\beta^2)*1/3*Y0, see Eq. 20 of Nozawa2006.
We also assume the contribution to $\langle y \rangle$ generated by the CMB dipole, $\langle y \rangle_{\rm dipole} \simeq 2.6\times 10^{-7}$, is subtracted using external measurements of the dipole amplitude~\cite{Chluba-Sunyaev2004}.

We now briefly summarize the model used to calculate Eq.~(\ref{eq.meantSZ}), with explicit details given in Ref.~\cite{Hilletal2014} and B15.  We assume the \emph{WMAP9}+eCMB+BAO+$H_0$ maximum-likelihood cosmological parameters (e.g., $\sigma_8 = 0.817$ and $\Omega_m = 0.282$)~\cite{Hinshawetal2013} and the halo mass function of Ref.~\cite{Tinkeretal2010}.  We use the ICM electron pressure profile fitting function from Ref.~\cite{Battagliaetal2012}, which is extracted from cosmological hydrodynamics simulations~\cite{Battagliaetal2010}.  This pressure profile matches a wide range of recent tSZ and X-ray observations~(e.g.,~\cite{Sunetal2011,Planck2013stack,Hajianetal2013,Hill-Spergel2014,Grecoetal2015}) and fully specifies the ICM electron pressure as a function of cluster mass, redshift, and cluster-centric distance.  Finally, to calculate the relativistic corrections, we use the $T_{\rm e}(M,z)$ relation from Ref.~\cite{Arnaudetal2005}, with a $+20$\% correction applied to the masses derived from X-ray data in that work, to account for deviations from hydrostatic equilibrium in the ICM~\cite{Planck2015clusters}.

The sky-averaged tSZ signal also receives contributions from electrons in the IGM and during reionization.  Our reionization model is described in Ref.~\cite{Battagliaetal2013} and B15.  The IGM and reionization contributions are subdominant to the ICM signal by more than an order of magnitude.  They are approximated well by the non-relativistic tSZ spectrum due to the electrons' low temperature ($k_{\rm B} T_{\rm e} \lesssim$ few eV), and thus are fully characterized by the Compton-$y$ parameter.  We add these contributions to that from the ICM to obtain the total sky-averaged tSZ signal.  We verify the accuracy of our analytic calculations by comparing to numerical simulations, finding that the predicted $\langle y \rangle$ values agree to within $2$\% (see B15).

\begin{figure}
\centering
\includegraphics[width=0.5\textwidth]{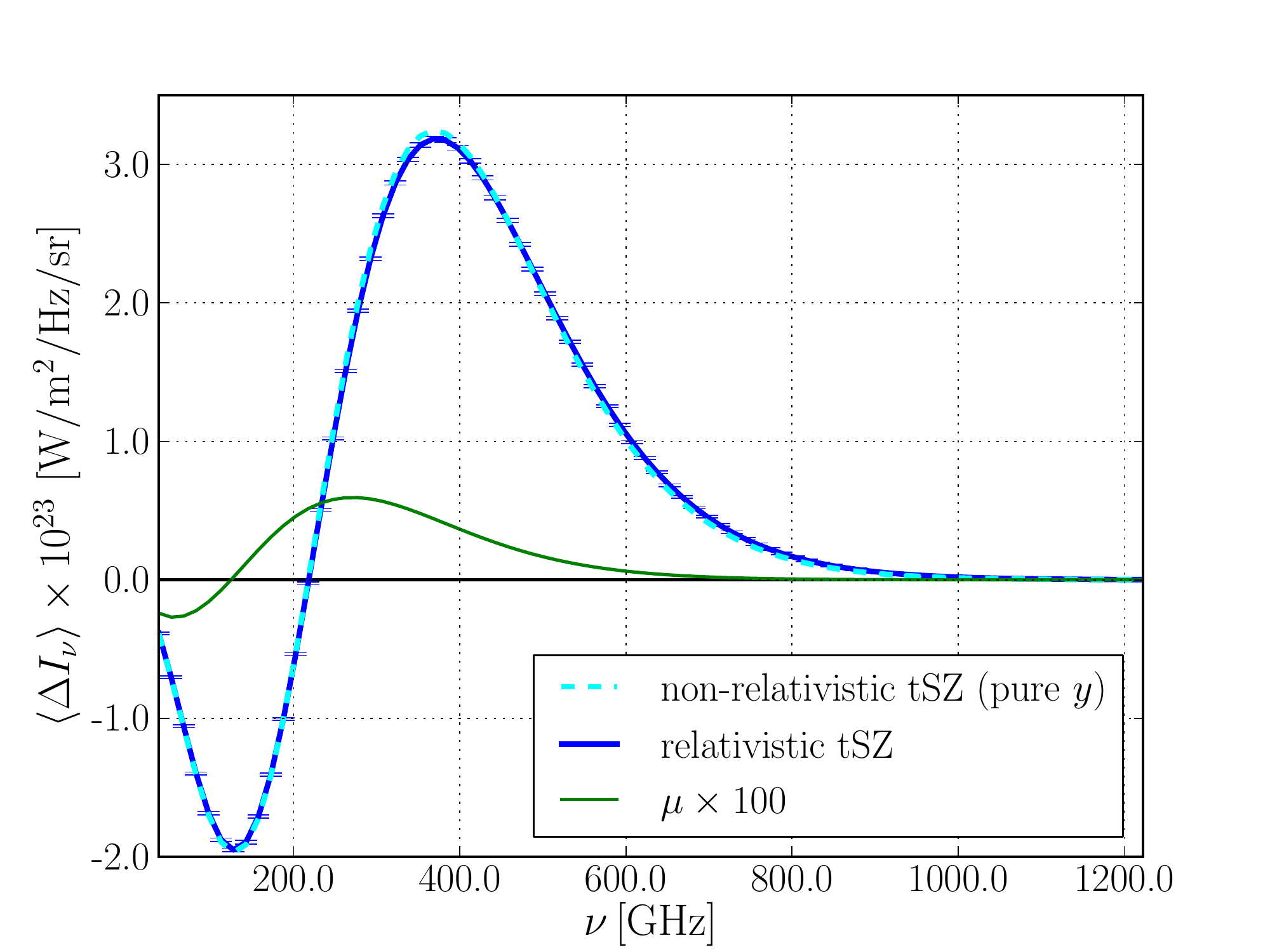}
\caption{The mean tSZ signal of the universe.  The dashed cyan and solid blue curves show non-relativistic and relativistic calculations, respectively.  The signal is dominated by hot, free electrons in galaxy groups and clusters.  Error bars are shown only on the relativistic curve for clarity, and include the \emph{PIXIE} instrumental noise, component separation noise, and cosmic variance (CV).  \emph{PIXIE} can detect the signal at $1470\sigma$ significance; CV reduces the effective signal-to-noise to $230\sigma$.  For comparison, the thin green curve shows the $\mu$ distortion (multiplied by $100$ to render it visible).
\label{fig.meantSZ}}
\end{figure}

Fig.~\ref{fig.meantSZ} shows the mean tSZ signal of the universe, for both the non-relativistic and relativistic cases, as well as the $\mu$ distortion signal for comparison.  %  Although the difference between the tSZ calculations is difficult to see, we show below that \emph{PIXIE} can measure the relativistic effects at high precision.
We show below that currently proposed experiments can measure the relativistic effects at high precision.
The non-relativistic results can be summarized fully by the Compton-$y$ parameter.  We find $\langle y \rangle_{\rm ICM} = 1.58 \times 10^{-6}$, $\langle y \rangle_{\rm IGM} = 8.9 \times 10^{-8}$, and $\langle y \rangle_{\rm reion} = 9.8 \times 10^{-8}$ for the contributions from the ICM, IGM, and reionization, respectively.  Note that $\langle y \rangle_{\rm ICM}$ depends sensitively on $\sigma_8$, the amplitude of matter density perturbations (going roughly as $\sigma_8^5$); if we assumed \emph{Planck} 2015 cosmological parameters~\cite{Planck2015params} instead of \emph{WMAP9}, the prediction would be $\approx 10$\% higher.  Regardless, the ICM contribution dominates over those from the IGM and reionization, although it may be possible to isolate the latter by masking the ICM using deep galaxy or cluster catalogs, or via cross-correlation techniques.  All $\langle y \rangle$ contributions are much larger in amplitude than the $\mu$ distortion signal.  In agreement with early estimates~\cite{Refregieretal2000}, the total $\langle y \rangle$ is roughly one order of magnitude below the \emph{COBE-FIRAS} bound.

\begin{figure}
\centering
\includegraphics[width=0.5\textwidth]{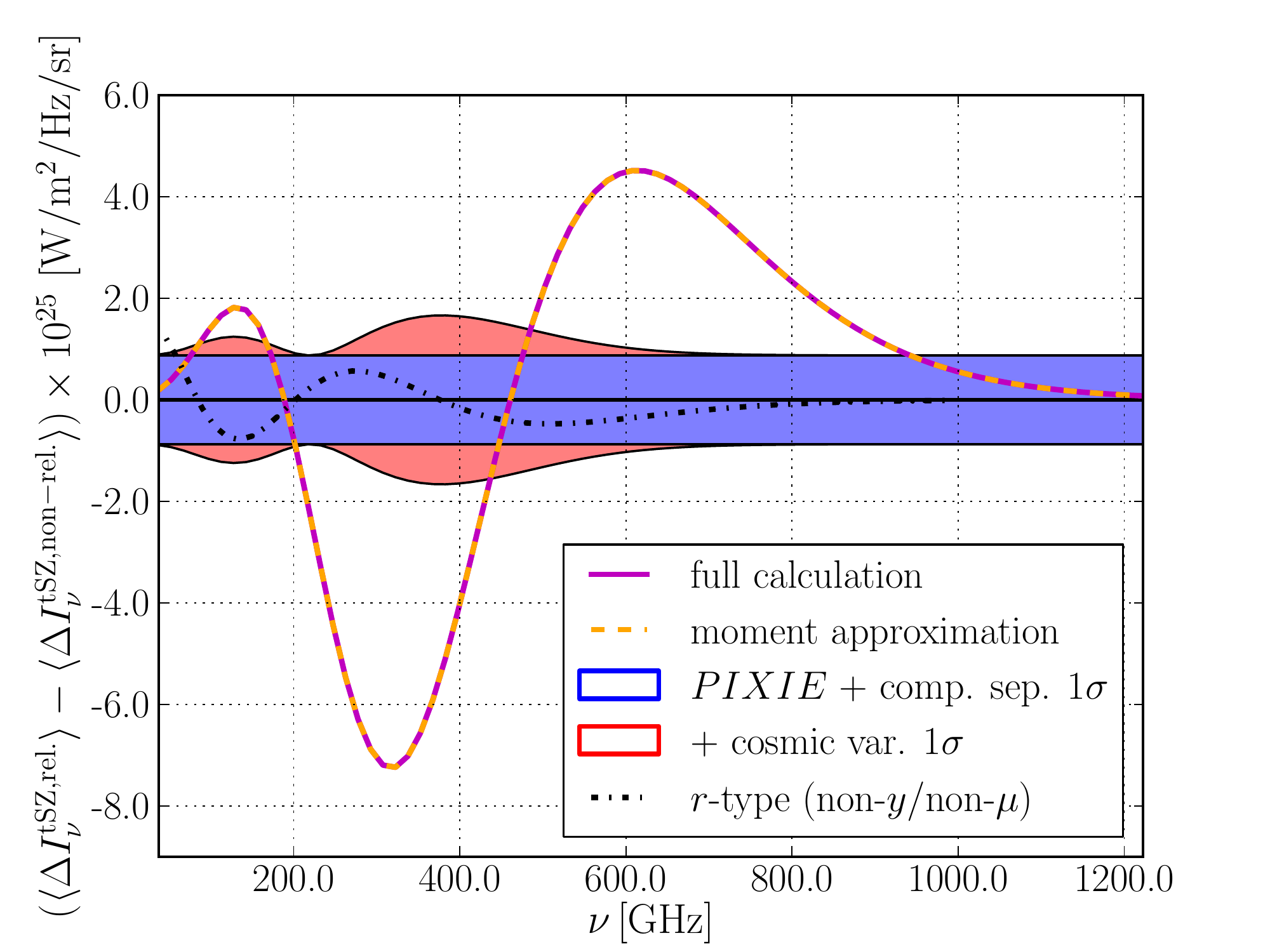}
\caption{Difference between relativistic and non-relativistic predictions for the mean tSZ signal of the universe.  %(the vertical axis values are two orders of magnitude smaller than in Fig.~\ref{fig.meantSZ}).
The solid magenta curve shows the difference between the tSZ predictions from Fig.~\ref{fig.meantSZ}.  The dashed orange curve shows an approximation %(but accurate) calculation in which relativistic corrections generated by an effective temperature $\langle k_{\rm B} T_{\rm e} \rangle_{\tau}$ are combined with a non-relativistic mean Compton-$y$, $\langle y \rangle_{\rm non-rel.}$.
based on moments of the optical depth-weighted ICM electron temperature distribution, which matches the full calculation to $\lesssim 0.1$\% precision.
The dash-dotted black curve shows the lowest-order ``residual'' (non-$y$/non-$\mu$) distortion~\cite{Chluba-Jeong2014}.  The relativistic tSZ signal is partially degenerate with this distortion, and thus makes it harder to access the $r$-type signal.
The shaded blue area shows the \emph{PIXIE} instrumental noise plus component separation noise, while the shaded red area shows the additional uncertainty from CV (note that important off-diagonal contributions are not shown).  \emph{PIXIE} can distinguish between the relativistic and non-relativistic predictions at $30\sigma$ significance.
\label{fig.relvsnonrel}}
\end{figure}

To emphasize the relativistic effects, Fig.~\ref{fig.relvsnonrel} shows the difference between the non-relativistic and relativistic predictions.  The fractional difference is $\approx 1$\% at $\nu \lesssim 500$ GHz, but is larger at higher frequencies (though the overall tSZ signal is smaller there).  
Fig.~\ref{fig.relvsnonrel} also shows the lowest-order ``residual'' ($r$-type) distortion (see Fig.~4 of~\cite{Chluba-Jeong2014}) assuming an amplitude $\mu_r = 10^{-6}$, which could be generated by (e.g.) decaying particles in the early universe (note that in our fiducial model $\mu_r = 0$; it is shown here only for reference).\footnote{Note that the $r$-type spectral shape was derived while neglecting relativistic tSZ effects (i.e., considering only pure $y$ and $\mu$)~\cite{Chluba-Jeong2014}.}  Fig.~\ref{fig.relvsnonrel} shows that the relativistic tSZ signal projects strongly onto the $r$-type distortion and thus renders its extraction more difficult.%  Thus, revised calculations of the residual distortion eigenmodes accessible to CMB spectral distortion experiments will be required.}

To interpret the relativistic tSZ signal, we consider a model based on moments of the optical depth-weighted ICM electron temperature distribution~\cite{ChlubaSwitzer2013}:
\ba
\label{eq.kTmomdef}
& & \langle (k_{\rm B} T_{\rm e})^n \rangle_{\tau} = \frac{1}{\langle \tau \rangle_{\rm ICM}} \int dz \frac{d^2 V}{dz d\Omega} \int dM \frac{dn}{dM} \times \nonumber \\
  & & \int d^2 \nhat \, \sigma_{\rm T} \int dl \, n_{\rm e}(\nhat, l, M, z) \left[ k_{\rm B} T_{\rm e}(\nhat, l, M, z) \right]^n \,,
\ea
where $\langle \tau \rangle_{\rm ICM}$ is the mean Thomson optical depth in groups and clusters:
\ba
\label{eq.tauICM}
\langle \tau \rangle_{\rm ICM} = & & \int dz \frac{d^2 V}{dzd\Omega} \int dM \frac{dn}{dM} \times \nonumber \\
  & & \int d^2 \nhat \, \sigma_{\rm T} \int dl \, n_{\rm e}(\nhat, l, M, z) \,.
\ea
We make the approximation that only electrons in groups and clusters are hot enough to require a relativistic treatment, i.e., the IGM and reionization contributions are fully characterized by $\langle y \rangle_{\rm IGM} + \langle y \rangle_{\rm reion}$ (which cannot be individually distinguished).  We implement the moment formalism for the ICM electrons following Ref.~\cite{ChlubaSwitzer2013}, considering moments up to $n=4$.

The model thus contains six parameters: $\langle (k_{\rm B} T_{\rm e})^n \rangle_{\tau}$ for $n=1$ to $4$, $\langle y \rangle_{\rm IGM} + \langle y \rangle_{\rm reion}$, and the total $\langle y \rangle$, which is interchangeable with $\langle \tau \rangle_{\rm ICM}$ via $\langle \tau \rangle_{\rm ICM} = (\langle y \rangle - \langle y \rangle_{\rm IGM} - \langle y \rangle_{\rm reion}) m_{\rm e} c^2/\langle k_{\rm B} T_{\rm e} \rangle_{\tau}$.  The fiducial values are $\langle (k_{\rm B} T_{\rm e}) \rangle_{\tau} = 0.208$ keV, $\langle (k_{\rm B} T_{\rm e})^2 \rangle_{\tau} = 0.299$ keV$^2$, $\langle (k_{\rm B} T_{\rm e})^3 \rangle_{\tau} = 0.892$ keV$^3$, $\langle (k_{\rm B} T_{\rm e})^4 \rangle_{\tau} = 4.02$ keV$^4$, $\langle y \rangle_{\rm IGM} + \langle y \rangle_{\rm reion} = 1.87 \times 10^{-7}$, and $\langle y \rangle = 1.77 \times 10^{-6}$ (equivalently, $\langle \tau \rangle_{\rm ICM} = 3.89 \times 10^{-3}$).  The prediction of this model for the relativistic deviation is shown in Fig.~\ref{fig.relvsnonrel}.  It agrees with the full calculation to $\lesssim 0.1$\% precision over the entire frequency range.  Thus, a measurement of the mean tSZ signal can be robustly interpreted in terms of these parameters.  B15 presents a detailed breakdown of the mass and redshift contributions to these quantities.

\emph{Prospects for Upcoming Experiments---}
\emph{PIXIE} is a proposed Explorer-class mission designed to measure the absolute intensity and linear polarization of the sky at $1.6$ degree resolution in 400 linearly-spaced frequency channels from 30 GHz to 6 THz~\cite{PIXIE2011}.  \emph{PIXIE}'s unprecedented spectral coverage and resolution will allow for exquisite component separation, yielding (low-angular resolution) maps of the tSZ signal, Galactic dust and synchrotron emission, and the cosmic infrared background, as well as precise constraints on primordial tensor fluctuations from the inflationary epoch.  We consider 80 frequency channels of width $\Delta \nu_{P} = 15$ GHz ranging from 30 GHz to 1230 GHz, with top-hat bandpasses. The tSZ signal is negligible at higher frequencies, but we assume the \emph{PIXIE} data in these channels will be used to measure dust foregrounds with sufficient precision to subtract them to the level of the instrumental noise at the lower frequencies~(see Sec.\ 3.2 of Ref.~\cite{PIXIE2011}).  The instrumental noise in each channel is $\Delta I_{P} = 5 \times 10^{-26} \,\, {\rm W \, m^{-2} \, Hz^{-1} \, sr^{-1}}$, which we take to be uncorrelated between channels~\cite{PIXIE2011,Chluba-Jeong2014}.  To account for noise arising from component separation, we model the sky-averaged signal using the terms in Eq.~(\ref{eq.DeltaIdef}).  In the Fisher matrix approximation, we find that marginalizing over the $\Delta T$ and $\mu$ components yields an effective total noise on the tSZ contribution of $\Delta I_{P_{\rm tSZ}} = 8.8 \times 10^{-26} \,\, {\rm W \, m^{-2} \, Hz^{-1} \, sr^{-1}}$, implying a raw detection significance of $1470\sigma$ for the sky-averaged tSZ signal shown in Fig.~\ref{fig.meantSZ}.    %As we will see below, the \emph{PIXIE} noise is so low that cosmic variance dominates the uncertainties on the measured mean tSZ signal.  Thus, even if the instrumental noise or foreground-subtraction noise turns out to be substantially higher than expected (even by a factor of $5$), our qualitative conclusions will remain unchanged.

In addition to the diagonal contribution to the channel-channel covariance matrix from instrumental noise, there is a (non-diagonal) contribution from cosmic variance:
\be
\label{eq.CV}
{\rm Cov}^{\nu \nu'}_{\rm CV} = \frac{C_{\ell = 0}^{\nu \nu'}}{4\pi\fsky} \,,
\ee
where $C_{\ell = 0}^{\nu \nu'}$ is the zero mode of the relativistic tSZ cross-power spectrum at frequencies $\nu$ and $\nu'$, and $\fsky$ is the observed sky fraction.  We derive this result in B15.  We use the exact, full-sky results for the tSZ power spectrum from Ref.~\cite{Hill-Pajer2013} to compute $C_{\ell = 0}^{\nu \nu'}$, accounting for relativistic effects using the same model as in our mean tSZ calculation.  We assume $\fsky = 0.75$ for \emph{PIXIE}, which is reasonable given its spectral coverage and ability to remove foregrounds.  The full covariance matrix is
\be
\label{eq.Cov}
{\rm Cov}^{\nu \nu'} = \Delta I_{P_{\rm tSZ}}^2 \delta^{\nu \nu'} + {\rm Cov}^{\nu \nu'}_{\rm CV} \,,
\ee
where $\delta^{\nu \nu'}$ is the Kronecker $\delta$-function.  The CV term contains non-negligible off-diagonal contributions, since nearly the same population of groups and clusters is responsible for the signal at different frequencies.

Including the CV contribution to the covariance matrix in Eq.~(\ref{eq.Cov}), the effective significance of the sky-averaged tSZ measurement decreases to $230\sigma$, which is still extremely promising.  The ultimate CV limit for a full-sky, zero-noise experiment is $\sim 1000\sigma$, depending on how much information in the high-frequency relativistic tail can be accessed.  It would be possible to decrease the CV by masking massive, low-redshift clusters that contribute significantly to $C_{\ell = 0}^{\nu \nu'}$ but not as significantly to the mean tSZ signal, similar to low-$\ell$ measurements of the tSZ power spectrum~\cite{Shawetal2009,Hill-Pajer2013}.%  However, this masking would complicate the theoretical interpretation.  Given the high signal-to-noise already predicted, we do not consider masking further.

Fig.~\ref{fig.meantSZ} shows error bars computed from the square root of the diagonal elements of the covariance matrix, including CV.  The error bars are difficult to see, but are more visible in Fig.~\ref{fig.relvsnonrel}, where we show the effective \emph{PIXIE} noise (including component separation degradation) and the full errors including CV (note that the off-diagonal CV contributions are important, but not shown on the plot).  Using the full covariance matrix, we find that \emph{PIXIE} can distinguish between the relativistic and non-relativistic mean tSZ predictions at $30\sigma$ significance.

We use the Fisher matrix formalism to forecast \emph{PIXIE} constraints on the parameters of the moment-based model presented in Fig.~\ref{fig.relvsnonrel}.  Degeneracies prevent meaningful constraints from being placed on most of the parameters at \emph{PIXIE}'s noise level, but if only $\langle y \rangle$ and $\langle k_{\rm B} T_{\rm e} \rangle_{\tau}$ are allowed to vary, we find marginalized $1\sigma$ uncertainties of $\sigma_{\langle y \rangle} = 8.3 \times 10^{-9}$ and $\sigma_{\langle k_{\rm B} T_{\rm e} \rangle_{\tau}} = 7.6 \times 10^{-3}$ keV.  These values correspond to fractional uncertainties of $0.5$\% and $4$\%, respectively.  \emph{PIXIE} will thus improve the existing \emph{COBE-FIRAS} bound on $\langle y \rangle$ by three orders of magnitude, while simultaneously measuring the mean optical depth-weighted ICM electron temperature at high precision.  It may be possible to further improve these forecasts by combining with constraints on $\langle \tau \rangle_{\rm ICM}$ from patchy $\tau$ estimators applied to CMB polarization maps~\cite{Dvorkin-Smith2009}.  Finally, note that the inclusion of $\langle k_{\rm B} T_{\rm e} \rangle_{\tau}$ degrades the constraint on $\mu$ by $\approx 10$\% compared to a forecast considering only the non-relativistic tSZ signal ($\mu$ has been marginalized over in the constraints quoted above).
%  In the future, additional information on $\langle \tau \rangle_{\rm ICM}$ from 21 cm experiments, CO intensity mapping, and fluctuations of the low-frequency free-free distortions may become available.

\emph{Outlook---}
Spectral distortions of the CMB contain a vast and complementary set of cosmological information to that contained in the temperature fluctuation power spectrum and other spatial statistics.  Amongst this information lies the mean tSZ signal of the universe, which can be detected at a raw significance of nearly $1500\sigma$ with \emph{PIXIE}, yielding a precise constraint on the total thermal energy in electrons in the observable universe. %high-precision measurements of the mean Compton-$y$ parameter of the universe and the mean optical depth-weighted ICM electron temperature.  These quantities %--- or, more precisely, the full measured mean tSZ distortion $\Delta I_{\nu}^{\rm tSZ}$ ---
%represent a key ``integral constraint'' on the thermodynamic properties of gas in the universe, similar to the constraint imposed on models of reionization by measurements of the optical depth to reionization, $\tau_{\rm CMB}$.  Any realistic model of galaxy formation must satisfy the constraints on $\langle y \rangle$ and $\langle k_{\rm B} T_{\rm e} \rangle_{\tau}$, which depend on the global injection of energy into baryons over cosmic history.  Current galaxy formation simulations rely crucially on such feedback to obtain realistic galaxy populations~(e.g.,~\cite{Vogelsberger2014,Hopkins2014,Schaye2015}), and will thus be significantly affected by the \emph{PIXIE} constraints.

In B15, we translate our forecasts into direct constraints on parameters describing physical models of the ionized gas.  %in the universe, including combinations of the \emph{PIXIE} constraints with those from the tSZ power spectrum and cross-correlations with gravitational lensing maps.  Of particular note is the constraint on the amplitude of the cluster mass -- temperature scaling relation, which is obtained through the relativistic effects shown in Figs.~\ref{fig.meantSZ} and~\ref{fig.relvsnonrel}.  Combining this zero-point calibration with upcoming ICM temperature measurements from \emph{eROSITA} will yield a powerful data set for cluster cosmology and represents a key synergy between \emph{PIXIE} and upcoming X-ray missions.  
In addition, the relativistic effects considered here lead to mixing between the low-redshift tSZ signal and the primordial $\mu$ and ``residual'' ($r$-type) spectral distortions, which must be accounted for in future analyses (see Fig.~\ref{fig.relvsnonrel}).  Overall, a measurement of the mean tSZ signal of the universe represents an important step forward in our understanding of the thermodynamics of ionized gas in the universe.

{\small \emph{Acknowledgments.} We are grateful to Al Kogut for useful conversations.  This work was partially supported by a Junior Fellow award from the Simons Foundation to JCH. NB acknowledges support from the Lyman Spitzer Fellowship.  JC is supported by the Royal Society as a Royal Society University Research Fellow at the University of Cambridge, U.K.  JCH, SF, ES, and DNS acknowledge support from NSF grant AST1311756 and NASA grant NNX12AG72G.}

\end{document}